# Spectral-filtering-induced phase transition in passively mode-locked fiber lasers

OUMAIMA OUGRIGE,[1, *] FLORENT BESSIN,[1] CHARLES CIRET,[1] HERVE LEBLOND,[1] AND FRANÇOIS SANCHEZ[1]

[1] *Univ Angers, LPHIA, SFR MATRIX, F-49000 Angers, France*
*\**oumaima.ougrige@univ-angers.fr*

**We investigate the effect of a tunable spectral filter on the dynamics of a passively mode-locked fiber laser in the anomalous dispersion regime. The results show that noise-like pulse emission evolves toward bound-state regime as the filter bandwidth is reduced. Thanks to the Shannon entropy applied to dispersive Fourier transform signals, it is demonstrated that the system undergoes a phase transition.**

Passive-mode-locked fiber lasers are well known for generating ultrashort pulses ranging from picoseconds to a few femtoseconds, making them highly valuable in fields such as telecommunications, biomedical imaging, and material processing [1,2]. These ultrashort pulses arise from a balance among dispersion, nonlinear effects, gain and losses [3] characterizing a non-conservative system [4]. Such systems are well known for their wide range of dynamics and complex behavior which have attracted significant attention in recent decades. Indeed, apart from the regular single-pulse mode-locking regime, numerous regimes involving multiple pulses per cavity round-trip have been observed or theoretically predicted [5,10].

Several methods based on the optical Kerr effect such as nonlinear polarization evolution (NPE) and the nonlinear amplifying loop mirror (NALM) are commonly used to achieve passive mode-locking [6]. These techniques enable regimes such as harmonic mode-locking, multistability with respect to pump power, pulse splitting, pulse fragmentation and bound states of two or more pulses [7,11]. These regimes can be achieved by either a simple rotation of the intra-cavity polarization controllers or by adjusting the pump power. The dynamics of these multi-pulsing regimes have been extensively explored [7,8,11], revealing a rich and complex landscape of pulse behavior.

Recently, several studies have focused on generating a large packet of bound, identical pulses with a fixed phase relationship per cavity roundtrip, referred to as a soliton crystal [7,12,13]. This concept was also previously explored by Mitschke in a conservative passive ring cavity [14,15]. The first experimental observations of a soliton crystal in fiber lasers were achieved through careful adjustment of the polarization controllers under slightly anomalous cavity dispersion [12]. It was then shown that a soliton crystal could be generated regardless of the exact mode-locking mechanism [13]. However, the precise experimental conditions required for the formation of these structures have yet to be clearly identified.

In this paper, we study a passive fiber mode-locked laser and demonstrate that a bound state containing many solitons can be achieved through spectral filtering. This concept is grounded in two key scientific results. First, in a conservative passive cavity containing an ensemble of solitons, studies have shown that a spectral band-pass filter can induce a transition from a disordered state to an ordered one [16]. Second, in a fiber laser cavity, both spectral gain and band-pass spectral filtering have been investigated. Haboucha et al [17] first reported a theoretical approach in the normal dispersion regime, demonstrating that reducing spectral filtering induces multi-pulsing. This was later confirmed in the case of a fiber laser passively mode-locked with a real saturable absorber [18]. Subsequent numerical simulations in the normal dispersion regime [19] demonstrated that filter characteristics can influence laser output, leading to multi-pulse, bound-state, single-pulse or noise-like pulses (NLP). Recent simulations also suggest that the spectral filter shape affects the fiber laser behavior [20]. Similar studies in the anomalous dispersion regime [21,22] have shown that spectral filtering can trigger multi-pulsing instability. In recent works, it has been experimentally shown that moderate spectral filtering can generate NLP with broad spectral bandwidth in an erbium-doped fiber laser operating in the anomalous dispersion regime [23]. Generally, in most experiments, wide bandpass filters tend to favor the emergence of NLP which can be viewed as a highly disordered state.

Precisely, in this work we aim to demonstrate that such a disordered NLP state can be transitioned to an ordered bound state by narrowing the spectral filter bandwidth. To achieve this, we incorporate a tunable super-Gaussian filter into our cavity and control the filter bandwidth to realize multiple bound soliton packets per cavity roundtrip. Analogous to phase transitions in thermodynamics, capturing, quantifying and characterizing the phase transition of the soliton states as the filter bandwidth is tuned requires a state function that exhibits sudden and significant changes in value during the transition. For this purpose, we employed the Shannon Entropy (SE), also known as information entropy, as the state function. Introduced in 1948 by C. E. Shannon, this powerful tool from information theory provides a standard metric for quantifying the system's degree of disorder or randomness, based entirely on the data generated by its source [24]. Such function has been used in various applications, from characterizing the optical spatial coherence of a light beam [25], to the soliton dynamics and evolution of unstable soliton clusters and Bose-Einstein condensate [26].

We considered applying Shannon entropy to temporal traces. However, this approach would require recording the motion of each individual soliton, an extremely challenging task. Such measurements demand either an oscilloscope and a photodiode with hundreds of gigahertz of bandwidth or a time lens. Even with these tools, only a short temporal window can be captured. Given these limitations, we opted for an alternative approach. Instead, for a more practical and effective analysis of the phase transition, we implemented a real-time spectroscopy technique in our setup known as Time Stretched Dispersive Fourier Transform (DFT). This technique allows us to capture the spectrum of each individual soliton packet in real-time [27,28], from which we can extract the Spectral Shannon Entropy (SSE). The SSE quantifies the distribution of spectral energy and, consequently, measures the disorder in a signal's frequency spectrum. SSE is mathematically computed as:

$$H = -\sum_{i=1}^{N} p(\omega_i) \ln(p(\omega_i))$$

where $p(\omega_i)$ is the amount of power contained in the spectral mode at angular frequency $\omega_i$, normalized by the total power in the spectrum, and $N$ is the number of frequency bins in the spectrum. Low SSE indicates that a signal's energy is concentrated in a few frequency components, implying regularity, predictability in the soliton packet. In contrast, high SSE indicates that the energy is distributed across many frequency components, reflecting complexity, randomness, or noise. The combination of these two techniques allows for real-time monitoring of the disorder within each individual soliton packet and enables tracking of phase transitions, indicated by a significant drop in the SSE.

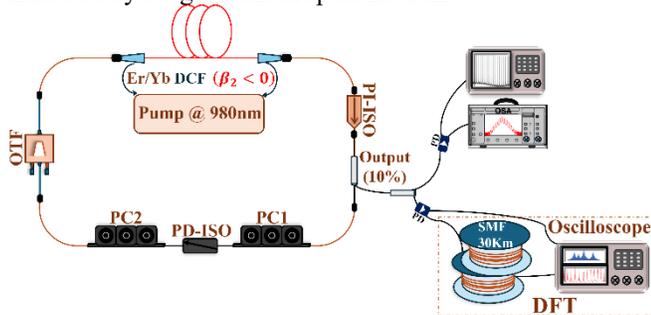

**Fig. 1.** Experimental setup

The all-fibered erbium-doped laser we built is presented in Fig 1. It is a unidirectional ring cavity incorporating an erbium-doped double-clad fiber (DCF) amplifier. A polarization-independent isolator (PI-ISO) is placed after the amplifier to ensure unidirectional operation and negates any parasitic reflections. Mode-locking is achieved through nonlinear polarization evolution (NPE), with nonlinear losses controlled using a polarization-dependent isolator (PD-ISO) placed between two polarization controllers (PC1 and PC2). Additionally, a super-gaussian optical tunable filter (OTF) is introduced into the cavity to trigger the generation of bound states. This filter is specially designed to enable precise adjustment of its lower and upper edges, tunable from 1520 nm to 1560 nm and 1530 nm to 1570 nm, respectively. However, the resulting spectral bandwidth is limited to a maximum value of 18 nm. The total cavity length is approximately 37 m, corresponding to a fundamental cavity frequency of 5.6 MHz and to a round-trip time of 179 ns, and the average dispersion of the cavity is about -0.0204 ps²/m at 1550 nm.

The laser output is extracted through the 10% port of a 90/10 fiber coupler placed between the PI-ISO and the PC1. The output intensity is detected with a fast photodiode (12 GHz bandwidth) and visualized with both a real-time oscilloscope (Agilent, Infinium DS08134B, bandwidth 13 GHz, 40 GSa/s) and an RF spectrum analyzer (Rhode & Schwarz, FSP 13, bandwidth 13 GHz). Additionally, we recorded the laser output spectral properties with (i) an optical spectrum analyzer (OSA) which capture on a slow-scale variations of the signal spectrum by averaging over a relatively long time window, and with (ii) DFT that can record single shot spectra in real time, which consists in a 30 km of single mode fiber with a total dispersion of -660 ps² that stretches the soliton packets, and a high-speed photodiode.

Before discussing the results, it is essential to outline the key initial conditions of the experiment. First, the laser is configured to generate bound states by arbitrarily centering the OTF with a narrow spectral width, typically a few nanometers. Once this state is achieved, the polarization controllers and pump power remain fixed throughout the entire experiment. In this configuration, when the filter is fully opened, the laser operates in the noise-like pulses regime. Then, its center wavelength ($\lambda_F$) and bandwidth ($\Delta\lambda_F$) are gradually adjusted to explore the potential for crystallizing or freezing an ensemble of solitons. It should be noted that the initial and final central wavelengths of the OTF can differ, as the tunability is realized through the independent adjustment of its edges.

In the first step, the intracavity filter was fully opened with its upper and lower edge wavelengths set to 1544 nm and 1562 nm, respectively, enabling the generation of noise like pulse regime (NLP). We then gradually reduced the tunable filter bandwidth, while simultaneously adjusting the polarization controllers and the pumping power until reaching a strongly modulated optical spectrum. This resulted in the transition from a large optical NLP spectrum to the modulated spectrum shown in Fig. 2. The spectrum is centered at 1559 nm and exhibits a strong modulation with a spectral period of 0.80 nm, revealing the existence of bound solitons with a temporal separation of about 10 ps. This bound-state regime is realized with a tunable filter bandwidth of about 4 nm. The cavity parameters are then fixed to their actual values, except the OTF. Our procedure ensures that the cavity can support bound states.

In a second step, our objective was to verify whether the bound state regime could be induced under the same conditions by maintaining the positions of the polarization controllers and the pump power fixed while varying only the OTF parameters: bandwidth ($\Delta\lambda_F$) and central wavelength ($\lambda_F$).

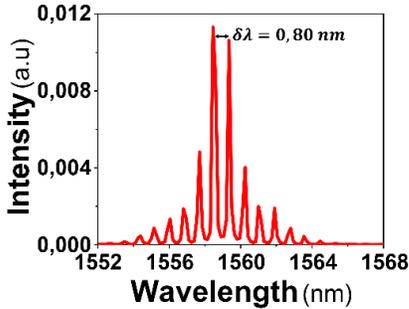

**Fig. 2.** Bound state spectrum when the filter is centered at $\lambda_F$ =1559 nm.

Initially, $\Delta\lambda_F$ was set to its maximum (18 nm), which consistently generated an NLP regime regardless of the value of $\lambda_F$. Here, we present results for the case where the initial central wavelength ($\lambda_F$) was set to 1550 nm. Figure 3 summarizes the experimental results, showing the evolution of the optical spectrum as a function of the optical filter bandwidth. When the filter is fully opened (18 nm bandwidth), the spectral width at the 10 dB level reaches approximately 127 nm, as shown by the black curve in Fig. 3(a). As $\Delta\lambda_F$ was progressively narrowed, the NLP regime persisted, as seen by the pink to purple curves in Fig. 3(a) and the spectral width varies by no more than 5 nm. When $\Delta\lambda_F$ was reduced below 6 nm, the bound state regime emerged, as illustrated by the blue curve in Fig. 3(a). Figure 3(b) provides a zoomed-in view that clearly shows the spectral modulation characteristics of bound solitons. As we can see, the system moves from a disordered pulse regime to a more stable state with organized bound state solitons as the filter bandwidth is reduced. This transition from the NLP regime to bound state is abrupt, resembling a phase transition process. This assumption is firstly supported by the fact that the resulting state (bound solitons) exhibits higher symmetry compared to the initial state (NLP) as shown in Fig. 3(a) (as evident from the purple curve and dark and light blue ones). From Fig. 3(b), the spectral period $\delta\lambda$ was determined to be approximately 0.72 nm, corresponding to a temporal separation between bound solitons within a packet of about 11 ps. Additionally, the recorded temporal trace showing the envelop of bound soliton packets depicted in Fig. 4(a), indicates that the laser operates in harmonic mode-locking regime with bound state packets of small extent, where each packet has an average duration of 0.25 ns. This is confirmed by the RF spectrum displayed in Fig. 5, which exhibits a peak at about 320 MHz corresponding to the 57-th harmonic. The average packet duration along with $\delta\lambda$, allow us to estimate the average number of pulses inside each soliton packet which is approximately 23 pulses. Secondly, the differences in temporal properties between NLP and bound state solitons also support the argument for a phase transition. Notably, the duration of the NLP packet as shown in Fig. 4(b) (2.26 ns) is significantly greater than that of a bound state soliton packet (0.25 ns), which suggests a transition from a broader, more randomized energy distribution to a more structured and localized pulse arrangement, where solitons remain confined within short bursts, as shown in Fig. 4(a).

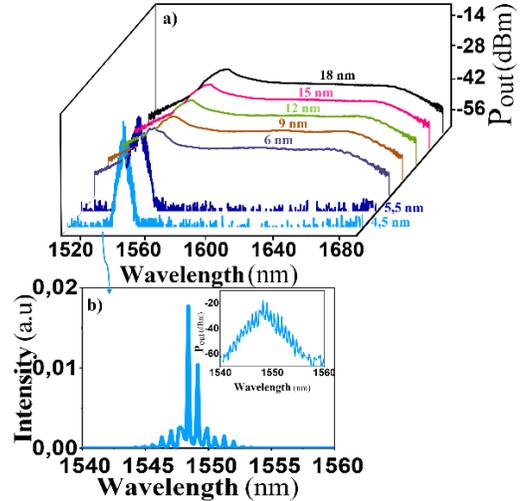

**Fig. 3.** Optical spectra. (a) The spectrum transitions when varying $\Delta\lambda_F$ from 18 nm to 4.5 nm. (b) The bound state regime when $\Delta\lambda_F$ is at 4.5 nm.

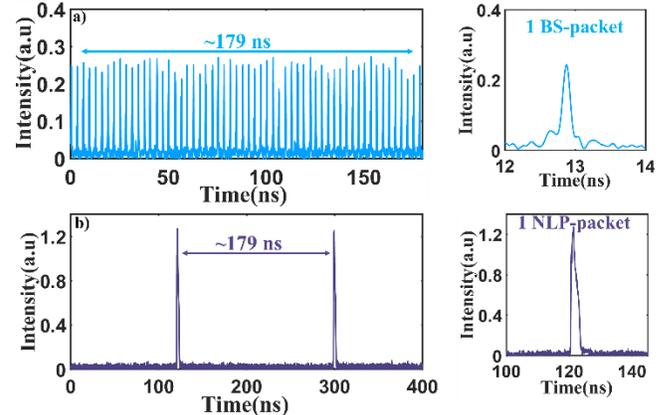

**Fig. 4.** (a) The temporal trace of the generated bound state (BS) regime at $\Delta\lambda_F$ = 4.5 nm. (b) The temporal trace of the generated NLP regime at $\Delta\lambda_F$ = 6 nm.

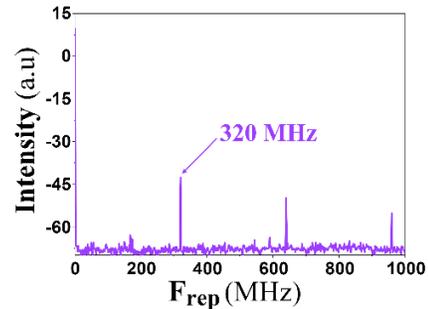

**Fig. 5.** RF spectrum showing harmonic corresponding at 320 MHz.

Moreover, to capture and quantify this phase transition from NLP to the bound state soliton regime, we recorded DFT traces for each of the six filter configurations as the $\Delta\lambda_F$ was tuned from 18 nm to 4 nm. These traces provide the real-time spectrum of each soliton packet. Figure 6(a) shows an example of a trace obtained in the NLP regime, where we observe that the packet is stretched by a factor of 3.

In contrast, Fig. 6(b) displays the DFT trace recorded when the bound state soliton regime is established. Notably, in the bound state soliton regime, each cavity round-trip yields the spectrum of all 57 soliton packets. We then calculate the average SSE from the DFT spectra over 287 cavity roundtrips recorded for each NLP regime, and similarly, for each of the 57 soliton packets in the bound-state regime. Figure 6(c) displays these results, illustrating the average SSE versus the filter bandwidth. In the NLP regimes, the average SSE remains relatively constant as the filter bandwidth decreases from 18 to 6 nm. However, when the filter is further reduced from 6 nm to 4 nm, a significant drop of the SSE from 9.4 to 5.4 is observed (well beyond the SSE measurement standard deviation, which remains below $1.5 \times 10^{-2}$). This drop suggests a shift from a highly disordered state to an ordered or partially ordered state, indicative of a phase transition.

Finally, to verify that our experimental measurements are not wavelength dependent, the experiments were conducted at $\lambda_F$=1560 nm, in addition to 1550 nm as previously presented, and at $\lambda_F$=1536 nm. The results were reproducible across all wavelengths, supporting the robustness of our findings.

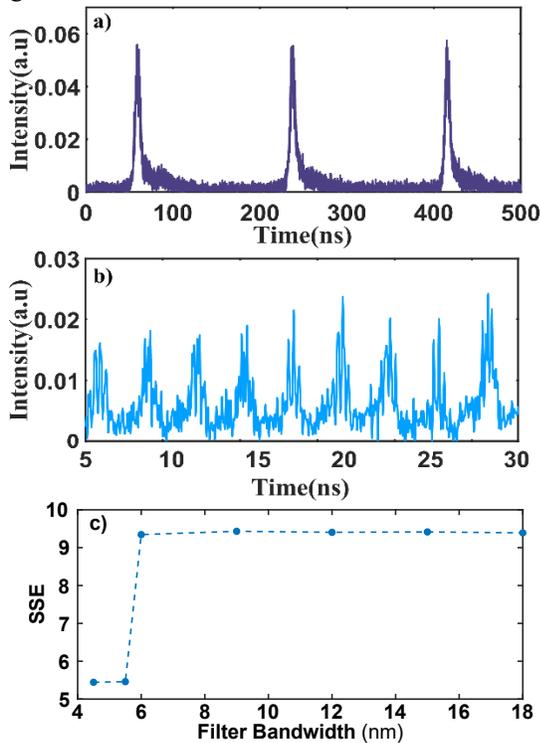

**Fig. 6.** (a) The DFT trace of the generated NLP regime at a $\Delta\lambda_F$ = 6 nm. (b) The DFT trace of the generated bound state regime at a $\Delta\lambda_F$ = 4.5 nm. (c) SSE versus the filter bandwidth.

A direct comparison of our results with those of Mitschke on conservative solitons [16] may not be straightforward, as our study focuses on dissipative solitons. However, our findings indicate that cooling or freezing effects may also occur in a dissipative system. We expect that closed solitons favor direct interaction, promoting the formation of bound structures.

In this study, we explored the role of intracavity spectral filtering in shaping the behavior of mode-locked fiber lasers. Our experiments revealed that progressively reducing filter bandwidth, the laser undergoes a transition from an NLP regime, a disordered state, to an ordered bound soliton state. This highlights the role of spectral filtering in driving a process akin to a phase transition within the laser cavity. To confirm the existence of a phase transition, it was necessary to define a state function and to point out a strong variation of this function versus the experimental control parameter (the filter bandwidth in our case). We have successfully used the Shannon entropy applied to the DFT signal to confirm this phase transition. Indeed, the SSE undergoes a significant drop as the system evolved from a disordered NLP regime to an ordered state composed of bound solitons. To ensure that the observed transition was not wavelength dependent, we repeated the experimental procedure with the filter set at shorter and longer wavelengths, confirming robustness and reproducibility of our findings.

In conclusion, our results open the way for thermodynamics studies of dissipative solitons in fiber lasers.

**Disclosures.** *"The authors declare no conflicts of interest."*